\documentclass[9pt,article,twoside]{rmaa-rho-class/rmac-rho}
\RMxAAtemplatetype{\RMxAC} 

\vol{61}
\pages{30-36}
\thisyear{2025}
\doi{\href{https://www.astroscu.unam.mx/RMxAC/RMxAC..XX-X}{https://www.astroscu.unam.mx/RMxAC/RMxAC..XX-X}}

\title{NOCTURNE. III. Unidentified variable emission in the nuclear regions of PKS 2153-69}

\author[1,2,3,$\dagger$]{M. ~Coloma ~Puga \orcidlink{0009-0001-4945-5781}}
\author[1]{M.~Berton \orcidlink{0000-0002-2653-1120}}
\author[1,4]{P.~Condò \orcidlink{0000-0001-6208-9109}}
\author[1,3,$\dagger$]{A. García-Pérez \orcidlink{0000-0002-9896-6430}}
\author[1]{A.~Jiménez-Gallardo \orcidlink{0000-0003-4413-7722}}
\author[6]{E.~Järvelä \orcidlink{0000-0001-9194-7168}}
\author[7,8]{A.~Lähteenmäki \orcidlink{0000-0002-0393-0647}}
\author[,9]{S.~Panda \orcidlink{0000-0002-5854-7426}\thanks{Gemini Science Fellow}}

\affil[1]{European Southern Observatory (ESO),Alonso de Córdova 3107, Casilla 19, Santiago 19001, Chile}
\affil[2]{INAF - Osservatorio Astrofisico di Torino, Via Osservatorio 20, I-10025 Pino Torinese, Italy}
\affil[3]{Dipartimento di Fisica, Universit\`a degli Studi di Torino, Via Pietro Giuria 1, 10125 (Torino), Italy}
\affil[4]{Dipartimento di Fisica, Università degli Studi di Roma “Tor Vergata”, via della Ricerca Scientifica 1, I-00133 Roma, Italy}
\affil[5]{Dipartimento di Fisica, “Sapienza” Università di Roma, Piazzale Aldo Moro 2, I-00185 Roma, Italy}
\affil[6]{Department of Physics and Astronomy, Texas Tech University, Box 41051, Lubbock, 79409-1051, TX, USA}
\affil[7]{Aalto University Metsähovi Radio Observatory, Metsähovintie 114, FI-02540, Kylmälä, Finland}
\affil[8]{Aalto University Department of Electronics and Nanoengineering, P.O. Box 15500, FI-00076 AALTO, Finland}
\affil[9]{International Gemini Observatory/NSF NOIRLab, Casilla 603, La Serena, Chile}
\affil[$\dagger$]{This study is part of project NOCTURNE}

\leadauthor{Coloma Puga et al.}
\smalltitle{LaTex Macro RMxAA}

\corres{Miguel Coloma Puga}
\email{miguel.colomapuga@inaf.it}

\received{April 16th, 2024}
\accepted{\today}

\license{Texto de la licencia aquí}

\setbool{rho-abstract}{true} 
\setbool{rho-resumen}{true} 

\begin{abstract}
    Historically, the study of the central regions of Type 1 AGN has been limited by the combination of the host galaxy spectrum with strong emission from the accretion disk and NLR/BLR, which prevented us from accurately probing the galactic and AGN properties in the central regions. Integral field spectroscopy allows us to correct for this effect and study both the unobscured cores of AGN host galaxies as well as the uncontaminated spectra of their central engines with unprecedented precision. Using MUSE WFM observations, in this work, we present a combined method for modelling and subtracting QSO light in type-1 AGN alongside results for one such source, PKS 2153-68 (z=0.028), both jetted and gamma-ray emitting. After separating the host galaxy and AGN spectra, we discuss the discovery of an unresolved and yet-to-be-identified high-velocity ($\sim$25000 km/s) short timescale ($\leq$1 yr) variable emission, unlike anything observed in other variable AGN.
\end{abstract}

\begin{resumen}
    Históricamente, el estudio de las regiones centrales de los AGN de Tipo 1 se ha visto limitado por la combinación del espectro de la galaxia anfitriona con la fuerte emisión proveniente del disco de acreción y NLR/BLR, lo cual nos impedía explorar las propiedades galácticas y del AGN en las regiones centrales. La espectroscopía de campo integral nos permite corregir estos efectos y estudiar tanto los núcleos de las galaxias anfitrionas de AGN como sus motores centrales con una precisión sin precedente. Haciendo uso de observaciones con MUSE WFM, en este trabajo presentamos un método combinado para el modelaje y sustracción de la luz del cuásar en AGN Tipo 1 junto a un estudio de una de estas fuentes, PKS 2153-68 (z=0.028), con presencia de un jet y emisión en rayos gamma. Tras separar la emisión de la galaxia anfitriona y el AGN, discutimos el descubrimiento de un complejo de emisión variable (escala temporal de $\sim$1 año) de alta velocidad ($\sim$25000 km/s) no resuelta y no identificada, disímil a cualquier emisión hasta ahora observada en AGN variables.
\end{resumen}

\begin{document}

\maketitle
\pagestyle{fancy}\thispagestyle{firststyle}


\section{INTRODUCTION}

\begin{figure}[ht]
    \includegraphics[width=\linewidth]{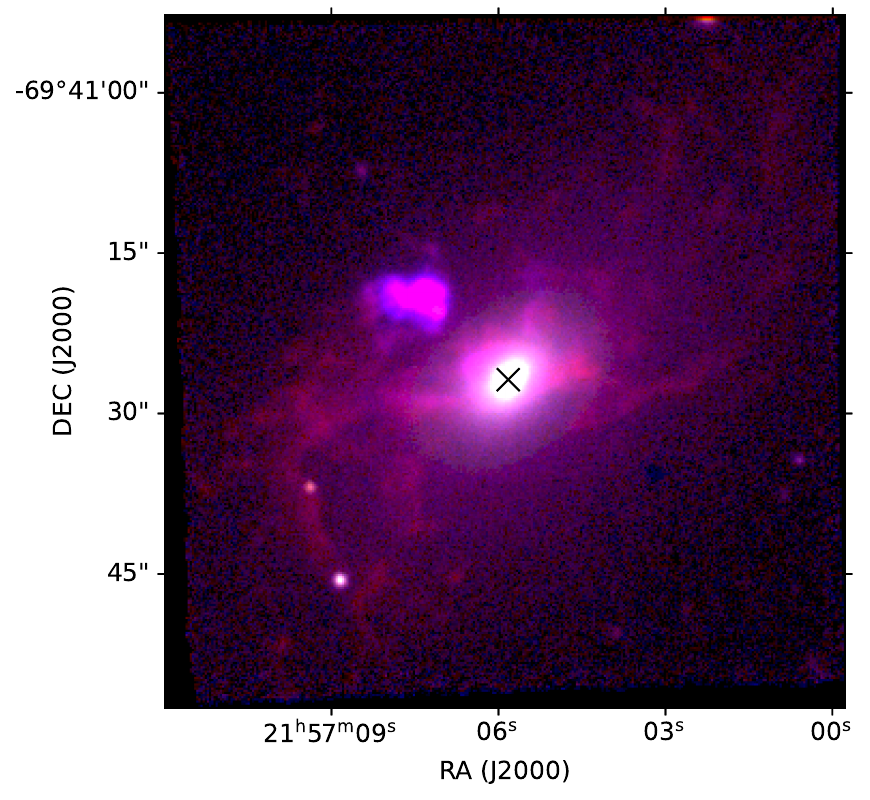}
    \caption{\justifying RGB image of PKS 2153-69. The blue band corresponds to [O III]$\lambda$5007, the green band corresponds to the stellar continuum, and the red band corresponds to H$\alpha$. The black cross marks the AGN position}
    \label{fig:rgb}
\end{figure}

Galaxies that host active galactic nuclei (AGN) are one of the most interesting laboratories in which to study the effects of secular processes on galactic evolution. Of most interest are the co-evolution of the central supermassive black hole (SMBH) and the galaxy itself \citep{Kormendy_2013}, and the feedback driven by AGN-related phenomena such as outflows and jets, which remains a necessity in cosmological simulations \citep{Dubois_2016}. There exists a variety of approaches to the characterization of the effects of AGN on their host galaxies, be it through large samples aimed at estimating the impact of feedback on cosmological scales or thorough studies of small samples of sources using more powerful instrumentation, intent on uncovering the possible consequences of single AGN feedback events, as well as under which conditions these phenomena are more effective \citep{2024harrison}. For the latter approach, being able to simultaneously study the properties of both the host galaxy and the SMBH is ideal.

The current paradigm of AGN unification \citep{Urry1995} posits that Type 1 sources are comprised of objects in which the central engine is viewed totally or partially face-on. As such, this should in principle enable an observer to study both the SMBH and its associated phenomena, as well as the host galaxy, down to the instrumental limits. However, separating one contribution from the other is a non-trivial problem, which in many cases requires a combination of spatial and spectral data in order to achieve a satisfactory solution. Within this context, we set out to study nearby type 1 AGN with confirmed presence of both radio jets and ionized outflows using integral field spectroscopy, as these sources would allow us to understand the interplay between the two different modes of feedback and the galactic properties, as well as their possible connection to SMBH properties. One such source is PKS 2153-69.

PKS 2153-69 (z=0.028) is a broad-line radio galaxy \citep{1993Tadhunter} hosted in a SA0 galaxy, which is also a low-luminosity $\gamma$-ray emitter identified by Fermi \citep{ajello2020}. X-ray observations using Chandra revealed an interaction between the radio jet and a highly ionized cloud, with the jet creating cavities within the hot CGM (\citealt{2005Ly,2012Worrall}). The viewing angle of the jet is about 10$^\circ$, at least before the impact with the high-ionization cloud \citep{2012Worrall}. The jet had also been observed and characterized using very large baseline interferometry (TANAMI, \citealt{2010ojha,2020angioni}. While an earlier low-resolution X-ray spectrum obtained by BeppoSAX did not show any iron emission lines \citep{2006Grandi}, more recent XMM-Newton observations suggested the presence of an ionized wind \citep{2019Mehdipour}.

\section{METHODS}
PKS 2153-69 (Fig \ref{fig:rgb}) was observed on two different nights with the Multi-Unit Spectroscopic Explorer (MUSE, \citealp{Bacon2010}) on the Very Large Telescope (VLT), the 17th and 20th of June of 2023, divided into four exposures of 634 s, with two exposures taken each night (Program I.D.: 111.24UJ.003, P.I.: Bian). The observations had seeing measurements of 0.74"$\pm$0.04" and 1.39"$\pm$0.07", respectively. The conditions on the second night of observations were noticeably worse, meaning we do not use those data when subtracting the QSO light or analyzing the properties of the central SMBH. As such, the effective total time on source for PKS 2153-69 was 1268 s. We used the European Southern Observatory (ESO) MUSE pipeline (version 2.8.7) to obtain a fully reduced and calibrated data cube \citep{Weilbacher20}.

Enabled by the high S/N of the observations and the low redshift of this source, we are able to follow the approach described in \citep{2013husemann,2014husemann}, which iteratively subtracts the galactic emission from the central regions, yielding separate AGN and host cubes. The first step in this process is to obtain a PSF model for the observations. We used the brightest available star in the field as a template for a point source. The PSF itself is modeled by fitting a 2D Gaussian to the spatial flux distribution of the star, iterating over the entirety of MUSE's wavelength range in 150\r{A} wide bands. This width ensures that the PSF is well-sampled across the spectrum while minimizing artifacts that can arise from noise peaks when using narrower bands. 

We then select a square region with a 5-pixel side centered on the AGN position, which contains both the QSO and host emission. The size of this region was selected as slightly larger than the seeing of the observations, which was equivalent to $\sim$4 px. A 2-pixel-wide concentric square was selected as representative of the host galaxy emission. The host contribution to the total emission in the nuclear regions can be estimated by masking the previously defined central region and fitting the remainder of the host galaxy using GALFIT \citep{2010peng}. This yields an estimate of the host contribution in the central region, which is then subtracted. The remaining "pure AGN" spectrum is convolved by the PSF and subtracted from the cube, and this process is iterated upon until a stable solution is reached. In our case, we reached residuals of $<1\%$ within the 4th iteration.

\section{RESULTS}

\subsection{QSO-host deconvolution}

\begin{figure}[ht]
    \includegraphics[width=\linewidth]{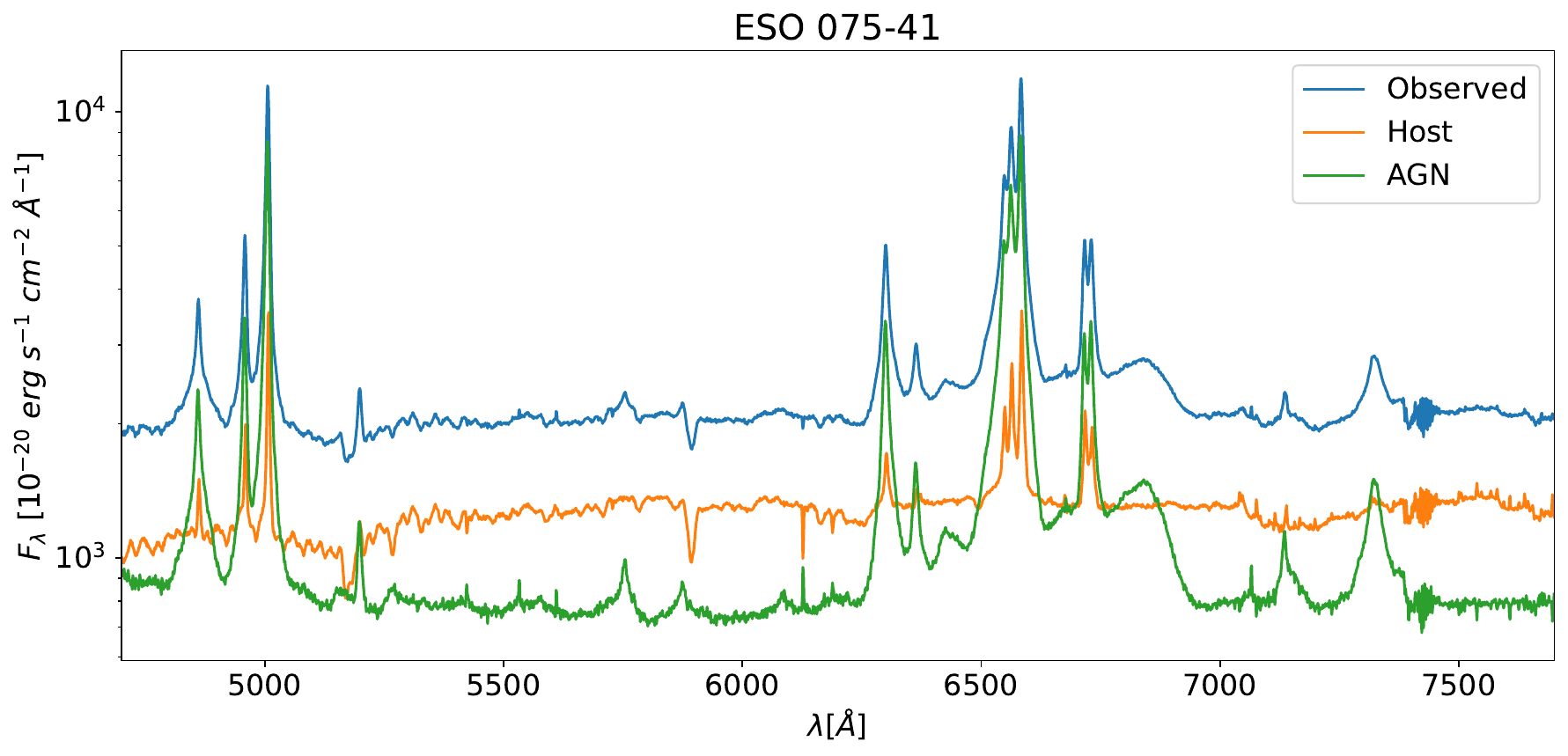}
    \caption{\justifying Deconvolved spectra of the AGN and host galaxy emission of PKS  2153-69. Wavelengths are in rest frame}
    \label{fig:sub}
\end{figure}

As mentioned previously, the deconvolution process yields two separate datacubes, one which includes the AGN spectrum convolved with the PSF, and another which is the result of subtracting this first cube from the original observation and, as such, should only include the host galaxy emission from both stars and gas. In Fig \ref{fig:sub} we present the observed spectrum within a circular aperture with a diameter equal to the PSF FWHM alongside the separate AGN and host galaxy spectra. The host galaxy spectrum consists of stellar continuum and absorption features as well as emission lines produced by the ionized gas, while the AGN spectrum includes a broader component in its lines and a flatter continuum. While all of these features were within expectations, we also observe an emission complex around the 6600\AA\ to 7100\AA\ range, which does not correspond to any previously known AGN emission and whose properties and nature are discussed in further detail in Sec. \ref{weird}.

\subsection{Variable nuclear emission}
\label{weird}

During the first look at the characteristics of the nuclear spectrum of PKS 2153-69, we noticed the presence of an extremely peculiar emission around the H$\alpha$/[N II] complex. Although the emission was blended with both the nearby lines and the stellar and AGN continua, which made a clear extraction difficult, it also became clear that the source of the emission was both point-like and either directly related to the SMBH or located in its immediate surroundings. No optical variability had been previously identified in this source, nor was it a known variable source in high-energy bands, meaning there was no monitoring of its spectra. While only an additional three archival optical spectra were available, they fortunately included epochs both before and after the MUSE observations. These observations include:

\begin{itemize}

    \item Combined optical and UV spectra from the ESO 3.6-m and the AAT 3.9-m telescopes (August 1978 to August 1986, \citealt{tadhunter1988})
    
    \item UVB, optical, and NIR spectra from XSHOOTER (June 2017, VLT, \citealt{koss2022})
    
    \item An optical spectrum from EFOSC2 (August 2024, NTT, P.I.: Jiménez-Gallardo, Program ID:113.26CQ.001)
    
\end{itemize}

\begin{figure*}[ht]
    \includegraphics[width=\linewidth]{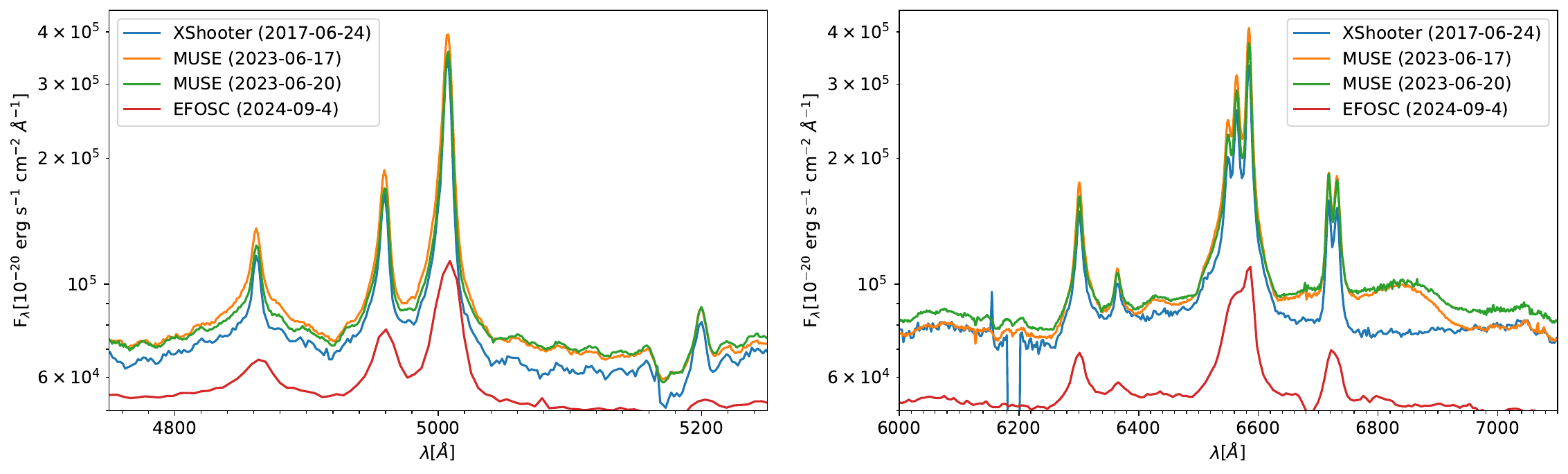}
    \caption{\justifying Comparison between the available nuclear optical spectra of PKS 2153-69. Both the MUSE and EFOSC2 spectra were extracted using an aperture the size of the PSF on each given night in order to maximize the signal-to-noise.}
    \label{fig:weird}
\end{figure*}

Given that the MUSE observations were performed in June of 2023, and that this emission is not observed in any of the other spectra (Fig. \ref{fig:weird}), we can both confirm its transient character and set an upper limit to the timescales of both its appearance ($\sim$6 yr) and disappearance ($\sim$1 yr).

Before continuing the analysis, we must first address the possibility of this emission stemming from a calibration issue, a hypothesis which we discard for two main reasons. Primarily, because there are two observations of this object on different nights (17th and 20th of June 2023), and while we do not use the observations from the second night on our analysis given the loss in spatial resolution, this emission is clearly visible in both cases. In addition, the spatial extension and morphology of the emission are equal to the PSF in both nights, and its location coincides exactly with the AGN position in both cases as well, which makes it all but impossible for this emission to be due to a systematic error.

A possible avenue through which we can study the variability of this AGN is its Eddington ratio. We use the broad components of the H$\beta$ line as a proxy for the virialized BLR gas around the SMBH and follow the method outlined in \citet{2025crepaldi}. Only the MUSE and XSHOOTER observations were used, as the other available spectra had too coarse a resolution and too low S/N to yield robust estimates of this parameter.

\begin{figure}[ht]
    \includegraphics[width=\linewidth]{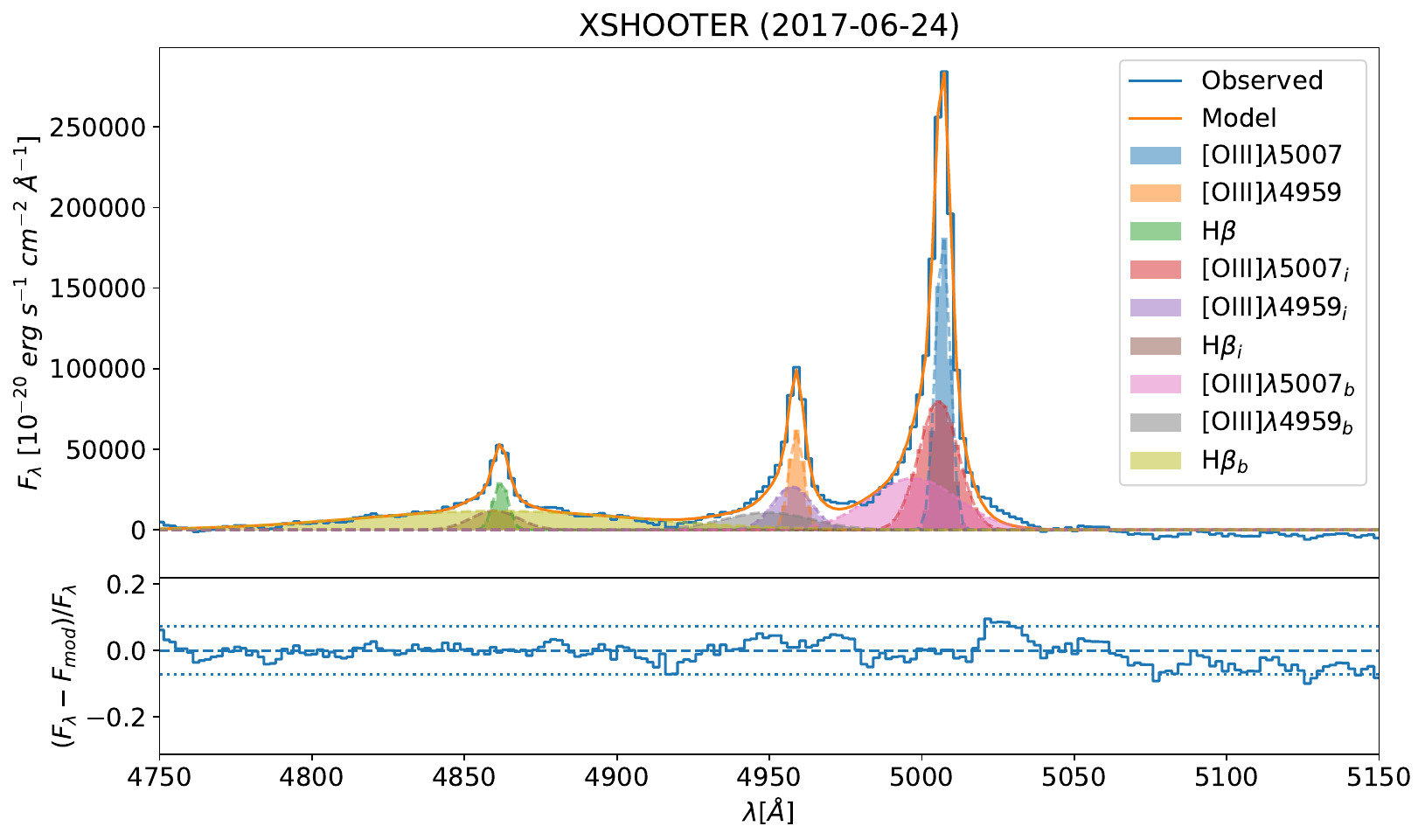}\\
    \includegraphics[width=\linewidth]{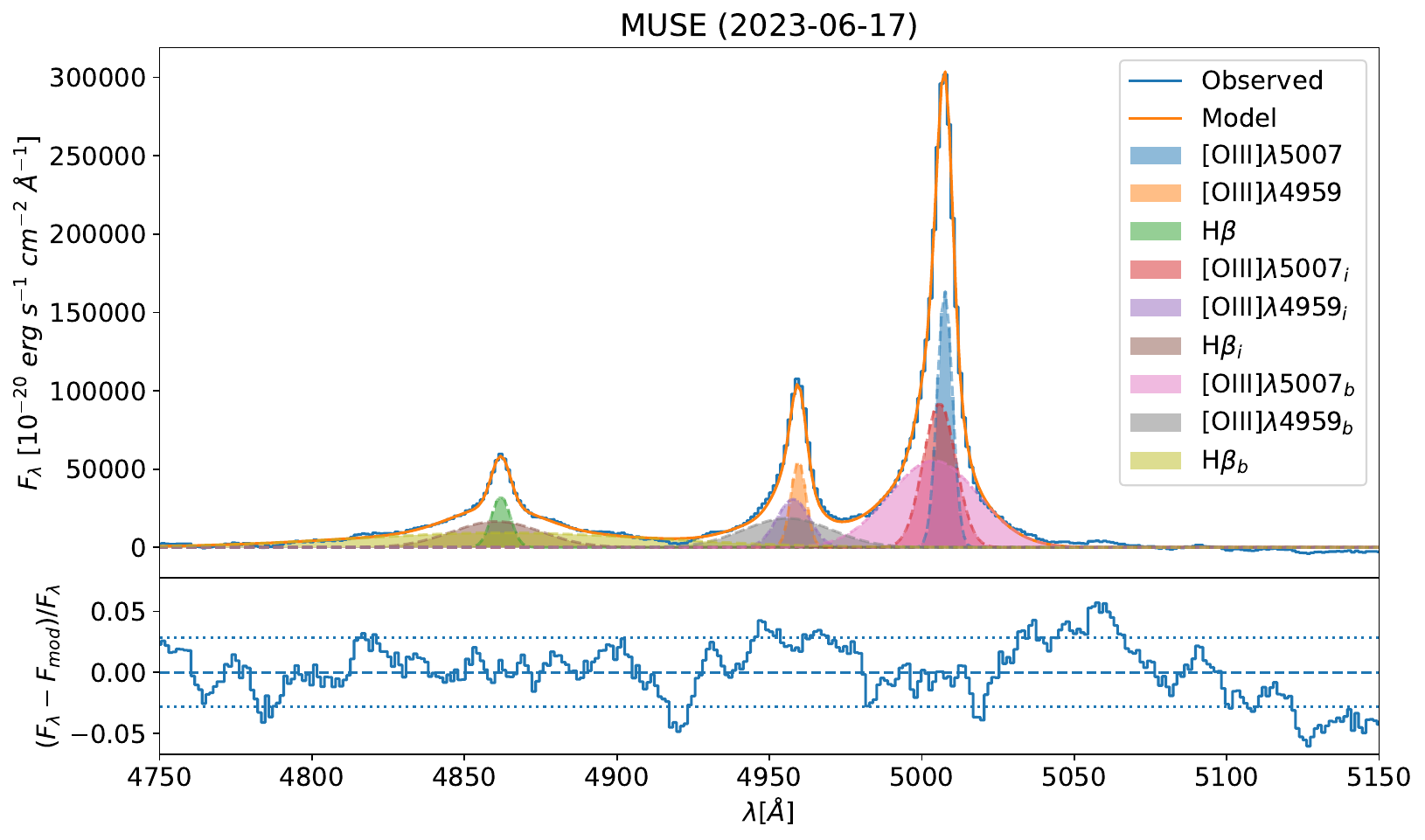}
    \caption{\justifying Spectral modeling of the H$\beta$/[O III] emission in PKS 2153-69 for two different epochs. At the bottom of each plot are the residuals expressed as the relative fraction of the flux. The dashed lines mark the measured noise levels.}
    \label{fig:smbh}
\end{figure}

Using a similar synthetic aperture for the MUSE data as that which was used in the XSHOOTER observations, the mass of the central black hole is estimated at log $\left({M_{BH}}/{M_{\odot}}\right)$=7.3$\pm$0.1 and we find the Eddington ratio remains almost perfectly constant, from an estimate of $\epsilon=0.008\pm0.005$ using the 2017 XSHOOTER observations to $\epsilon=0.009\pm0.003$ with MUSE. Indeed, the spectral profile of H$\beta$ and the resulting models are extremely similar between the two epochs (see Fig. \ref{fig:smbh}). Additional details on the calculations used for these estimates will be provided in Coloma Puga et al., in prep.

\section{DISCUSSION}


The identification of the nature and physical origin of this emission remains an issue. While the overall shape of this spectral feature somewhat resembles that of a relativistic disk-like BLR (see e.g. \citealt{bianchi2019}), being an extremely broad ($v\sim25000\, km\,s^{-1}$ between the two peaks) asymmetric emission close to a Balmer line, it deviates from this hypothesis in four ways:

\begin{itemize}
    \item There's another broad component identifiable on the nuclear spectrum in all epochs, with FWHM $\sim$4000 km s$^{-1}$. This component is both compact and present in both Balmer lines (as well as the rest of the ionized lines in the spectrum), ensuring its identification as BLR emission
    \item If the two observable peaks are the blue and red peaks of the relativistic BLR emission, its centroid is offset by $\sim$5000 km s$^{-1}$ with respect to the narrow nuclear H$\alpha$ emission; an offset of the BLR with respect to the NLR this large has never been observed in any other object.
    \item It does not appear around H$\beta$, whereas a regular relativistic BLR should appear around all Balmer lines.
    \item The bright peak is on the red side instead of the blue one, something highly atypical for relativistic BLR emission.
\end{itemize}

Although this last point could be explained through geometrical effects (e.g., an asymmetric accretion disk or one containing a spiral arm, see \citealt{eracleous2009}), the other three remain difficult to contend with. This observation either does not represent any type of currently known or well-defined AGN emission, or it is extremely affected by radiative transfer and absorption effects to the point where it resembles something else entirely. For instance, while the lack of any emission around H$\beta$ could plausibly be explained by an extreme amount of dust obscuration in the nucleus, it would not only require extinction orders of magnitude higher than that present in the rest of the galaxy, but the obscuring object would have to be perfectly aligned with the source of the emission.

Given the fact that this emission was not visible before the MUSE observations, and became too faint to be observed afterwards, we believe it is more likely for this variability to be driven by a short-timescale physical phenomenon rather than it being a long-time feature of this AGN that becomes visible when obscuration subsides. This hypothesis, however, is also not without demerit. There was no variability observed in any of the energy bands from the Fermi-LAT light curves around the time of our observations, and the only significant flaring visible in the entirety of the available light curve was in early 2014.

As it stands, further observations of this source, particularly in its tentative "flaring" state, will be required to fully ascertain the nature of its variability. Future facilities such as the Vera C. Rubin Telescope (formerly LSST) will provide multi-epoch high resolution spectra of transient sources such as this one, allowing for more detailed studies.


\renewcommand{\refname}{REFERENCES}
\bibliography{rmxac}


\footnotesize
\section*{ACKNOWLEDGEMENTS}
Based on observations collected at the European Southern Observatory under ESO programmes 111.24UJ, 113.26CQ, and 099.A-0403(B). This study is part of the project NOCTURNE\footnote{\href{https://www.ejarvela.space/nocturne/}{https://www.ejarvela.space/nocturne/}}, which stands for Narrow-line Seyfert 1 galaxies Over Cosmic Time: Unification, Reclassification, Nature, and Evolution. NOCTURNE aims to improve our overall understanding of the population of undermassive black holes, such as NLS1s, by embracing a panchromatic approach to investigating unevolved AGN. SP is supported by the international Gemini Observatory, a program of NSF NOIRLab, which is managed by the Association of Universities for Research in Astronomy (AURA) under a cooperative agreement with the U.S. National Science Foundation, on behalf of the Gemini partnership of Argentina, Brazil, Canada, Chile, the Republic of Korea, and the United States of America.

\end{document}